\documentclass{article}

     \PassOptionsToPackage{numbers, compress}{natbib}

     \usepackage[preprint]{neurips_2024}

\usepackage[utf8]{inputenc} 
\usepackage[T1]{fontenc}    
\usepackage{hyperref}       
\usepackage{url}            
\usepackage{booktabs}       
\usepackage{amsfonts}       
\usepackage{nicefrac}       
\usepackage{microtype}      
\usepackage{xcolor}         

\usepackage[pdftex]{graphicx}

\usepackage[whole]{bxcjkjatype}

\def\etal{{\it et al.}}
\def\ddG{{$\Delta \Delta G$}} 

\def\flexddg{{Flex ddG}}

\title{Active Learning for Energy-Based Antibody Optimization and Enhanced Screening}

\author{
Kairi Furui \quad\quad\quad Masahito Ohue$^{*}$\\
Institute of Science Tokyo\\
\texttt{furui@li.c.titech.ac.jp}, \ \  \texttt{ohue@c.titech.ac.jp} \\
$*$ Corresponding author
}

\begin{document}

\maketitle

\begin{abstract}
    Accurate prediction and optimization of protein-protein binding affinity is crucial for therapeutic antibody development. Although machine learning-based prediction methods $\Delta\Delta G$ are suitable for large-scale mutant screening, they struggle to predict the effects of multiple mutations for targets without existing binders. Energy function-based methods, though more accurate, are time consuming and not ideal for large-scale screening. To address this, we propose an active learning workflow that efficiently trains a deep learning model to learn energy functions for specific targets, combining the advantages of both approaches. Our method integrates the RDE-Network deep learning model with Rosetta's energy function-based Flex ddG to efficiently explore mutants. In a case study targeting HER2-binding Trastuzumab mutants, our approach significantly improved the screening performance over random selection and demonstrated the ability to identify mutants with better binding properties without experimental $\Delta\Delta G$ data. This workflow advances computational antibody design by combining machine learning, physics-based computations, and active learning to achieve more efficient antibody development.
\end{abstract}

\section{Introduction}

Optimizing the binding affinity between proteins is an important aspect of drug development, including antibody engineering~\cite{kim2023computational,10.1093/bib/bbae307}. In recent years, deep learning-based {\it de novo} molecular design methods have been proposed \cite{shuai2023iglm,jin2022antibody,martinkus2024abdiffuser,shanehsazzadeh2023unlocking,luo2022antigen,ueki2024antibody}. However, approaches that utilize existing antibody information to design improved variants are also important, as {\it de novo} methods may not fully leverage the knowledge of existing antibodies~\cite{shan2022deep,zeng2024antibody,hoie2023antifold,tennenhouse2024computational}. This is particularly crucial when dealing with known antibodies that require improvements in desirable properties such as binding affinity, specificity, and developability.
In recent years, many mutation-induced binding free energy change (\ddG) prediction methods based on machine learning and deep learning have been proposed~\cite{shan2022deep,li2016mutabind,ZHANG2020100939,rodrigues2019mcsm,luo2023rotamer,Yu2024ddaffinity,liu2024predicting}, and their prediction accuracy has improved. However, in cases where there are insufficient data on existing antibodies, the prediction results may not be reliable.
In such situations, even if the predicted \ddG{} for mutants, especially multiple mutations, is optimized, overfitting may occur, making it difficult to propose improved sequences in practice.
In addition, traditional methods for calculating \ddG{} of mutants based on energy functions~\cite{stranges2013comparison,buss2018foldx,Barlowflexddg} are applicable even when sufficient \ddG{} information about known mutants is not available, and it is known that energy functions with structural sampling, such as Rosetta's energy function-based \ddG{} prediction method \flexddg, have high prediction accuracy. However, such methods that involve structural sampling require enormous computational costs compared to machine learning-based methods, making them unsuitable for large-scale screening of mutants.

In recent years, the pharmaceutical industry has shown an increasing interest in using a machine learning approach called active learning~\cite{settles2009active,reker2015active,reker2017active} to streamline time-consuming calculations such as compound docking and molecular simulations\cite{Wang2023-ht,konze2019reaction,loeffler2024optimal}.
Active learning is a method that iteratively selects the next data sample to be experimented based on a machine learning model called a surrogate model~\cite{settles2009active,reker2015active,reker2017active}, with the aim of efficiently conducting costly experiments such as experimental verification and molecular simulations.
Here, it is important to balance the exploration phase, which reduces uncertainty and selects data with high information content, with the exploitation phase, which discovers data with high activity or desirable properties, to select cost-effective data samples.
These methods are useful for exploring promising candidate samples even in the absence of extensive existing experimental data. By iteratively adding data, active learning can guide the optimization process more efficiently than conventional high-throughput screening methods.
For example, in the field of small molecule drug discovery, active learning combined with molecular simulations and docking simulations has already succeeded in improving screening performance without relying on experimental information~\cite{Wang2023-ht,konze2019reaction,loeffler2024optimal}. Similarly, several active learning approaches have been proposed in the field of antibody engineering~\cite{seo2022accelerating,lu2021k,gessner2024active}.
Seo \etal~\cite{seo2022accelerating} take an approach that combines a deep learning model based on the CTRL method with experimental validation by ELISA.
Lu \etal~\cite{lu2021k} propose a method called K-Nearest Robust Active Learning, KRAL, and apply it to epitope prediction in B cell data.
Gessner\etal~\cite{gessner2024active} discover antibody sequences with improved binding affinity based on relative binding free energies calculated using Bayesian optimization.
Such active learning approaches for experimental information or relative binding free energy calculations provide accurate predictions of binding affinity but require enormous costs, time, and difficult setups.
Also, to achieve sufficient performance with these approaches, a sufficient number of samples may be required, so a more convenient approach for faster pre-screening would be useful.
Furthermore, among these existing approaches, methods based on relative free energy perturbation calculations can be executed faster than experimental validation approaches if computational resources are available, but it is difficult to handle multiple mutations~\cite{zhu2022large}.

Therefore, we propose a new workflow that efficiently learns the scoring of the \flexddg{} method~\cite{Barlowflexddg} through active learning, in contrast to conventional methods, to construct a fast pre-screening model.
\flexddg{} achieves performance comparable to state-of-the-art models~\cite{lu2024alphafold3} but requires a relatively long computation time for structural sampling.
Our workflow guides the selection of promising antibody variants by combining the predictive performance of machine learning with the physics-based knowledge of \flexddg. In this study, we achieve this by incorporating both experimental \ddG{} values and computational binding information predictions through multitask learning based on the architecture of the RDE-Network developed by Luo \etal{}
This allows us to identify antibody sequences that may improve binding affinity while minimizing the need for large-scale experiments and enhancing the predictive performance of \flexddg's computational \ddG.
To demonstrate the effectiveness of our method, we optimize Trastuzumab (Herceptin\textsuperscript{\textregistered}), an antibody targeting the human epidermal growth factor receptor (HER2), and the results showed a significant improvement in screening performance compared to random selection. Furthermore, an interesting result was obtained that the binding classification performance could be improved even without experimental information on HER2 by using additional data based on \flexddg.
This study contributes to the growing field of efficient and effective computational antibody design by presenting a novel workflow combining machine learning, physics-based computation, and active learning.

\section{Materials and Methods}
\begin{figure}[tbp]
  \centering
  \includegraphics[width=\linewidth]{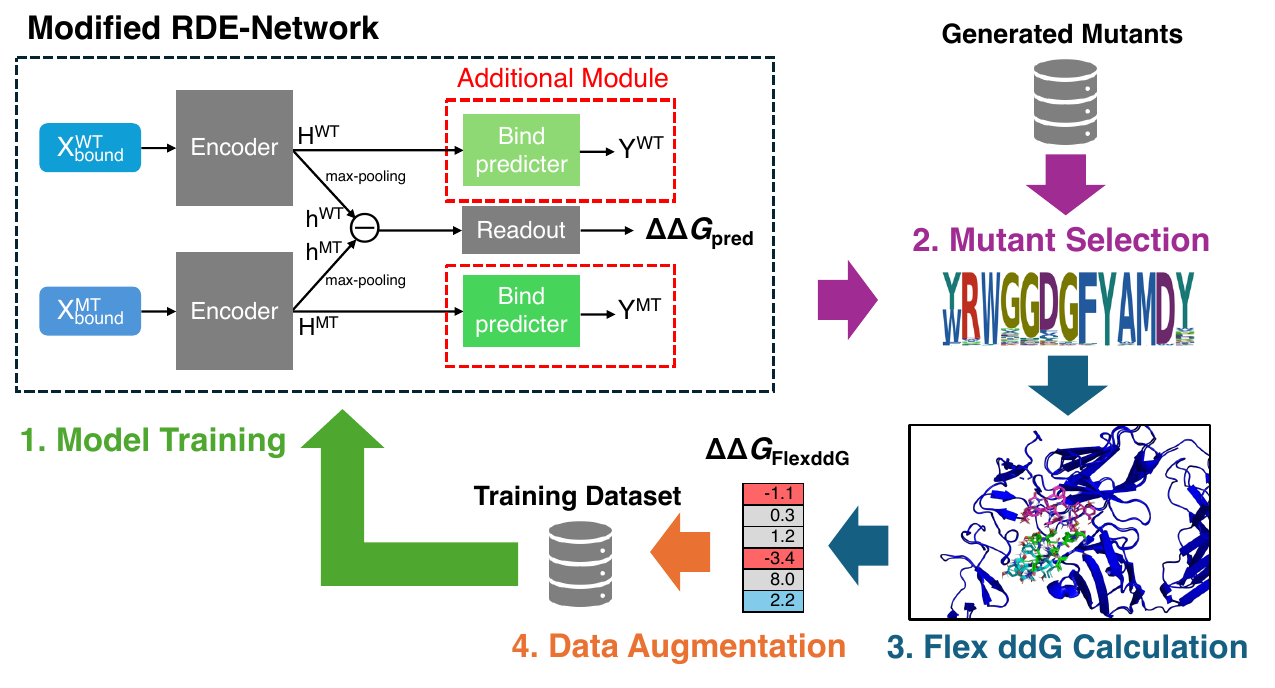}
  \caption{Overview of the proposed active learning workflow.}
  \label{fig:workflow}
\end{figure}

Figure~\ref{fig:workflow} shows an overview of the active learning workflow proposed in this study.

Trastuzumab, an antibody targeting the HER2 protein, was used as the antibody to be optimized, and the complementarity-determining regions (CDR-H) of the heavy chain were optimized. In this process, mutant data was sampled by randomly mutating the amino acid sequence of Trastuzumab's CDR-H (8 residues for CDR-H1, 8 residues for CDR-H2, and 13 residues for CDR-H3) to 19 amino acid residues excluding cysteine with a mutation probability of 0.2, generating 100,000 mutants (of which 98,567 were unique sequences).

Luo \etal's RDE-Network~\cite{luo2023rotamer} was used as the \flexddg{} surrogate model.
RDE-Network enables \ddG{} prediction performance by learning information on mutants included in the SKEMPI2 dataset~\cite{jankauskaite2019skempi} based on the predicted probability distribution, using a flow-based rotamer probability distribution model pretrained on crystal structures.
RDE-Network is one of the state-of-the-art models in terms of \ddG{} prediction performance for mutants.
Here, in addition to the \ddG{} prediction module existing in the original RDE-Network prediction model, a Binding Module that predicts whether to bind directly from the output layer of the wildtype and mutation RDE modules was added. In other words, multitask learning was performed using experimental \ddG{} and computational binding information.
For the Rotamer Density Estimator, pre-trained parameters were used and fixed during surrogate model training.
The learning parameters other than the newly added parts of the model were the same as those of Luo \etal, and the cross-entropy loss for binding was added with a weight of 0.5.
Unlike the data split of Luo \etal, the data split of the learning dataset was based on the Holdout information included in the SKEMPI2 dataset to ensure that PPBs similar to the PPBs of the validation set were not included in the training set.
Based on the holdout information, a relationship network of complexes included in the SKEMPI2 dataset was constructed, and the connected components of this network were treated as a single group. Based on these groups, a 3-fold group split was performed, using the fold containing HER2-Trastzumab data as validation data and the remaining folds as training data.
This excluded all \ddG{} information related to HER2 and antibodies from the training set prior to active learning.

In each cycle, 200 trastzumab mutants were selected and a total of 1200 were selected over 6 cycles of active learning.
In each cycle, only mutants that differed by 2 or more mutations from those selected so far, including previous cycles, were selected in descending order of the surrogate model's predictions, to ensure an appropriate balance between exploration and exploitation of the selected sequences.
The parameters of \flexddg{} were based on those of ~\cite{Barlowflexddg} with the number of structural ensembles set to $N=10$.
The \ddG{} calculated by \flexddg{} was added to the learning data in the next active learning cycle as binding data if it was less than 0 and as non-binding data if it was 2 or greater. Data with $0<\Delta \Delta G<2$ were excluded from the learning data.
Also, in the first cycle, the model was trained by the surrogate model without any HER2 mutation data.

As a test set for model evaluation, surface plasmon resonance (SPR) assay data of Trastuzumab mutants designed by Shanehsazzadeh \etal~\cite{shanehsazzadeh2023unlocking} was used.
At this time, \ddG{} was obtained from the $K_{\mathrm{D}}$ values of the SPR experiment, and only mutants with the same length of CDR-H3 as Trastuzumab were selected, which will be referred to as the Trastuzumab mutant dataset hereafter. The Trastuzumab mutant dataset includes 419 binding mutants and 846 non-binding mutants.

\section{Results and Discussion}
\begin{figure}[tbp]
  \centering
  \includegraphics[width=\linewidth]{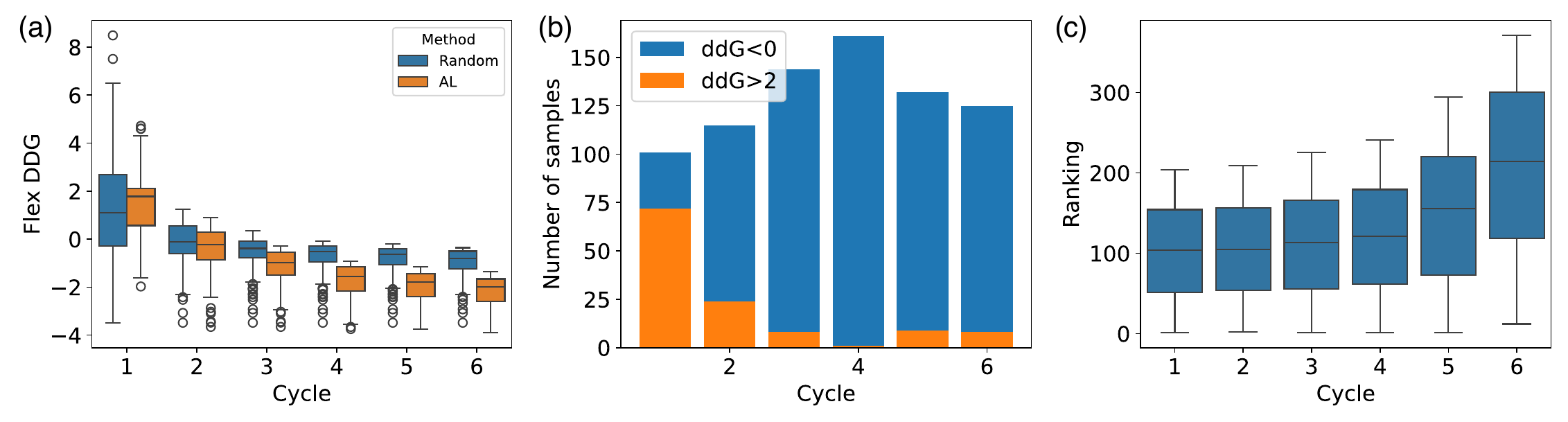}
  \caption{(a) Transition of the calculated top 200 \flexddg{} values of the selected mutants at each active learning cycle.
  (b) Number of selected mutants that bound and unbound based on \flexddg{} at each active learning cycle.}
  \label{fig:results-flex-calcs2}
\end{figure}

\begin{figure}[tbp]
  \centering
  \includegraphics[width=\linewidth]{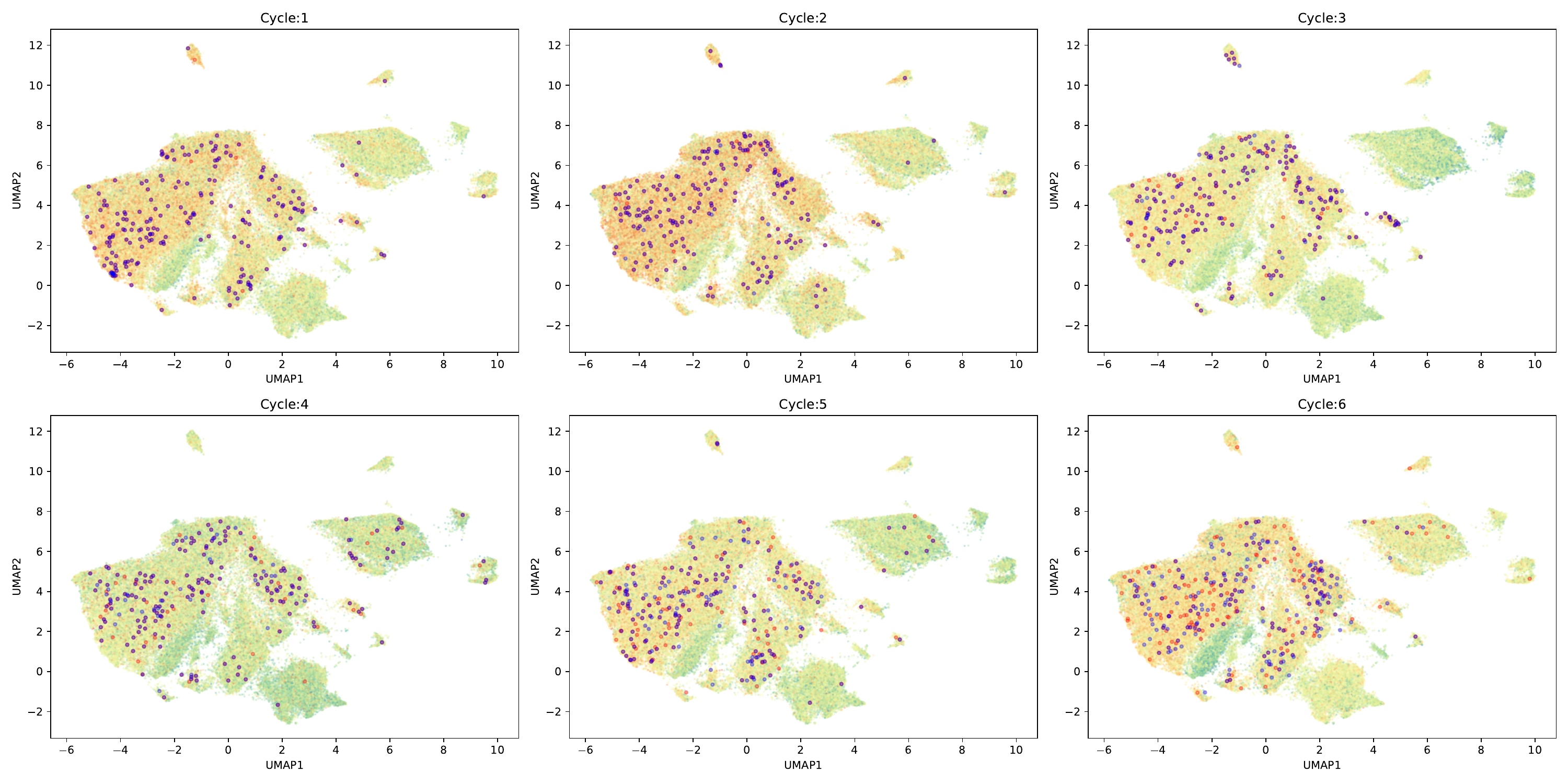}
  \caption{The H-chain sequences of the HER2 mutant dataset were embedded using AbLang~\cite{olsen2022ablang} and compressed to 2D using UMAP~\cite{mcinnes2018umap} for visualization. The red dots represent the selected mutants at each active learning cycle, and the blue dots represent the top 200 selections for that cycle. In the early cycles of active learning, the blue and red dots overlap, but in the later cycles, the red dots spread out more than the blue dots.}
  \label{fig:result-visualize}
\end{figure}

Figure~\ref{fig:results-flex-calcs2}(a) shows the transition of the distribution of the top 200 \flexddg{} values of the data selected in previous cycles.
For comparison, the results were compared with the case where 1200 data points were randomly selected from the data pool and evaluated by \flexddg.
The figure shows that the proposed method can discover Trastuzumab mutants with significantly lower \flexddg{} values compared to random selection in the final cycle.
In the early stages of learning, the top predictions of RDE-Network tended to select worse mutants than Random, but by using them for learning in subsequent cycles, the enrichment was improved.
Furthermore, in Figure~\ref{fig:results-flex-calcs2}(b), the number of selected mutants that bound and unbound based on \flexddg{} at each active learning cycle, it can be seen that in the initial learning, many unbound data were sampled due to inaccurate model predictions, but in the latter half of learning, almost no non-binding data were selected. Figure~\ref{fig:results-flex-calcs2}(c) shows the transition of the rank of the predicted \ddG{} of the selected mutants. In the early cycles, the selection is close to greedy selection, but in the latter half of learning, mutants with worse ranks are selected, which means that more diverse mutants were selected, i.e., more exploratory sampling was performed, due to the constraint of selecting only those that differed by 2 or more mutations from the previously selected ones. This can also be seen from the difference between greedy selection and actual selection in the embedded space in Figure~\ref{fig:result-visualize}. Thus, balancing exploitation and exploration is important for more efficient learning of active learning models.

\begin{figure}[tbp]
        \centering
        \includegraphics[width=\linewidth]{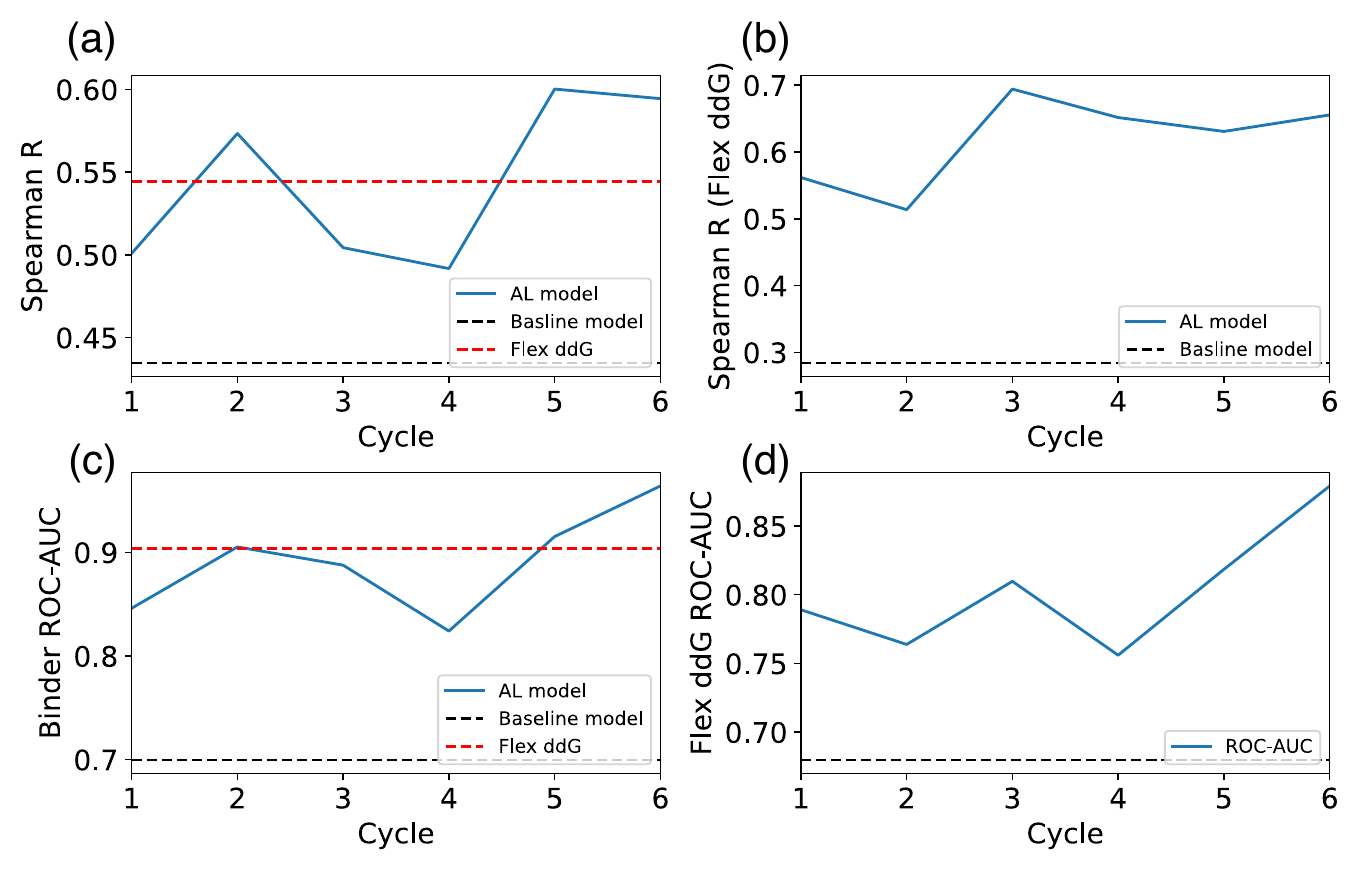}
        \caption{(a) Transition of Spearman correlation of the surrogate model for the Trastuzumab mutant dataset at each active learning cycle.
        (b) Transition of Spearman correlation on \flexddg{} of the surrogate model for the Trastuzumab mutant dataset at each active learning cycle. Each correlation was calculated only for those labeled as binders.
        (c) Transition of ROC-AUC scores for binding classification for the Trastuzumab mutant dataset at each active learning cycle.
        (d) Transition of ROC-AUC scores for binding classification on \flexddg{} for the Trastuzumab mutant dataset at each active learning cycle. The target variable was set as ``binder'' when $\Delta \Delta G_{\mathrm{\flexddg}}$ < 0.
        }
        \label{fig:results-all}
\end{figure}

Furthermore, Figures~\ref{fig:results-all}(a) to (d) show the Spearman correlation with experimental \ddG{} or \flexddg's \ddG{} and the ROC-AUC for experimental binding and \flexddg-based binding at each active learning cycle, respectively.
The Spearman correlation was calculated only for the binding mutants in the Trastuzumab mutant dataset, and for the binding ROC-AUC, 50 data points each were selected from the experimental binding data ($\Delta\Delta G_{\mathrm{\flexddg}}<0$) and randomly selected non-binding data for a total of 100 data points for comparison with \flexddg{} calculation.
From Figures~\ref{fig:results-all}(a) and \ref{fig:results-all}(b), it can be seen that the predictive performance of the active learning model slightly improved from the baseline and reached a level comparable to \flexddg. However, the correlation with experimental \ddG{} did not change even as the number of active learning cycles increased.
Also, Figures~\ref{fig:results-all}(c) and \ref{fig:results-all}(d) show a significant improvement in performance compared to the baseline model. Furthermore, in the latter half of active learning, the discriminative performance surpassed that of \flexddg.
This is an interesting result showing that the actual screening performance can be improved without using experimental information, just by adding the computational values of \flexddg{} to the learning.
From the above, the proposed method seems to be effective in improving the binding classification performance in particular.

\section{Conclusions}
In this study, we developed an active learning workflow for optimizing antibody sequences.
The proposed method successfully screened Trastuzumab variants against HER2 efficiently.
By utilizing active learning models as proposed in this study, we succeeded in efficiently screening by evaluating only a subset of the vast number of Trastuzumab variants against human HER2 using \flexddg, without calculating the energy of all sequences.
Furthermore, by combining the SKEMPI2 dataset and the computational energy information of variants' \flexddg, RDE-Network was able to more accurately discriminate the binding of human HER2 without learning the experimental \ddG{} information. This suggests that it may be useful for exploring variants with higher binding in the early stages of antibody design, where only a few weak binders are obtained, or for optimizing variants with better pharmacological properties.
In this study, RDE-Network was used as a surrogate model for active learning, but this framework is applicable to any model and can be useful for easily constructing models tailored to specific targets.
However, the active learning method proposed in this study is primarily for optimization and requires the preparation of complex structures of known binding ligands.
Furthermore, since the sequences evaluated in this study have not been experimentally validated, the effect may be limited if \flexddg{} calculations do not correlate well with the experimental \ddG, and it should be noted that in actual drug development phases, the developability of the selected variants should also be considered.
This study investigated offline-based optimization, but combining it with genetic algorithms, as in other cases, may cover a more diverse mutant space.
In the future, by combining active learning with more accurate affinity prediction methods that require even more extensive calculations, such as relative binding free energy perturbation calculations, we aim to construct models with higher binding affinity prediction accuracy for specific targets and improve screening accuracy.

\bibliographystyle{unsrt}
\bibliography{main.bib}

\begin{thebibliography}{10}

\bibitem{kim2023computational}
Jisun Kim, Matthew McFee, Qiao Fang, Osama Abdin, and Philip~M Kim.
\newblock Computational and artificial intelligence-based methods for antibody development.
\newblock {\em Trends Pharmacol. Sci.}, 44(3):175--189, 2023.

\bibitem{10.1093/bib/bbae307}
Sara Joubbi, Alessio Micheli, Paolo Milazzo, Giuseppe Maccari, Giorgio Ciano, Dario Cardamone, and Duccio Medini.
\newblock {Antibody design using deep learning: from sequence and structure design to affinity maturation}.
\newblock {\em Brief. Bioinform.}, 25(4):bbae307, 07 2024.

\bibitem{shuai2023iglm}
Richard~W Shuai, Jeffrey~A Ruffolo, and Jeffrey~J Gray.
\newblock {IgLM}: Infilling language modeling for antibody sequence design.
\newblock {\em Cell Syst.}, 14(11):979--989, 2023.

\bibitem{jin2022antibody}
Wengong Jin, Regina Barzilay, and Tommi Jaakkola.
\newblock Antibody-antigen docking and design via hierarchical equivariant refinement.
\newblock {\em arXiv preprint arXiv:2207.06616}, 2022.

\bibitem{martinkus2024abdiffuser}
Karolis Martinkus, Jan Ludwiczak, Wei-Ching Liang, Julien Lafrance-Vanasse, Isidro Hotzel, Arvind Rajpal, Yan Wu, Kyunghyun Cho, Richard Bonneau, Vladimir Gligorijevic, et~al.
\newblock {AbDiffuser}: full-atom generation of in-vitro functioning antibodies.
\newblock {\em Adv. Neural Inf. Process. Syst.}, 36, 2024.

\bibitem{shanehsazzadeh2023unlocking}
Amir Shanehsazzadeh, Sharrol Bachas, Matt McPartlon, George Kasun, John~M Sutton, Andrea~K Steiger, Richard Shuai, Christa Kohnert, Goran Rakocevic, Jahir~M Gutierrez, et~al.
\newblock Unlocking de novo antibody design with generative artificial intelligence.
\newblock {\em bioRxiv}, 2023.

\bibitem{luo2022antigen}
Shitong Luo, Yufeng Su, Xingang Peng, Sheng Wang, Jian Peng, and Jianzhu Ma.
\newblock Antigen-specific antibody design and optimization with diffusion-based generative models for protein structures.
\newblock {\em Adv. Neural Inf. Process. Syst.}, 35:9754--9767, 2022.

\bibitem{ueki2024antibody}
Takafumi Ueki and Masahito Ohue.
\newblock Antibody complementarity-determining region design using {AlphaFold2} and {DDG} predictor.
\newblock {\em J. Supercomput.}, 80:11989--12002, 2024.

\bibitem{shan2022deep}
Sisi Shan, Shitong Luo, Ziqing Yang, Junxian Hong, Yufeng Su, Fan Ding, Lili Fu, Chenyu Li, Peng Chen, Jianzhu Ma, et~al.
\newblock Deep learning guided optimization of human antibody against {SARS-CoV-2} variants with broad neutralization.
\newblock {\em Proc. Natl. Acad. Sci.}, 119(11):e2122954119, 2022.

\bibitem{zeng2024antibody}
Yimeng Zeng, Hunter Elliott, Phillip Maffettone, Peyton Greenside, Osbert Bastani, and Jacob~R Gardner.
\newblock Antibody design with constrained bayesian optimization.
\newblock In {\em ICLR 2024 Workshop on GEM}.

\bibitem{hoie2023antifold}
Magnus H{\o}ie, Alissa Hummer, Tobias Olsen, Morten Nielsen, and Charlotte Deane.
\newblock {AntiFold}: Improved antibody structure design using inverse folding.
\newblock In {\em NeurIPS 2023 GenBio Workshop}, 2023.

\bibitem{tennenhouse2024computational}
Ariel Tennenhouse, Lev Khmelnitsky, Razi Khalaila, Noa Yeshaya, Ashish Noronha, Moshit Lindzen, Emily~K Makowski, Ira Zaretsky, Yael~Fridmann Sirkis, Yael Galon-Wolfenson, et~al.
\newblock Computational optimization of antibody humanness and stability by systematic energy-based ranking.
\newblock {\em Nat. Biomed. Eng.}, 8(1):30--44, 2024.

\bibitem{li2016mutabind}
Minghui Li, Franco~L Simonetti, Alexander Goncearenco, and Anna~R Panchenko.
\newblock {MutaBind} estimates and interprets the effects of sequence variants on protein--protein interactions.
\newblock {\em Nucleic Acids Res.}, 44(W1):W494--W501, 2016.

\bibitem{ZHANG2020100939}
Ning Zhang, Yuting Chen, Haoyu Lu, Feiyang Zhao, Roberto~Vera Alvarez, Alexander Goncearenco, Anna~R. Panchenko, and Minghui Li.
\newblock {MutaBind2}: Predicting the impacts of single and multiple mutations on protein-protein interactions.
\newblock {\em iScience}, 23(3):100939, 2020.

\bibitem{rodrigues2019mcsm}
Carlos~HM Rodrigues, Yoochan Myung, Douglas~EV Pires, and David~B Ascher.
\newblock {mCSM-PPI2}: predicting the effects of mutations on protein--protein interactions.
\newblock {\em Nucleic Acids Res.}, 47(W1):W338--W344, 2019.

\bibitem{luo2023rotamer}
Shitong Luo, Yufeng Su, Zuofan Wu, Chenpeng Su, Jian Peng, and Jianzhu Ma.
\newblock Rotamer density estimator is an unsupervised learner of the effect of mutations on protein-protein interaction.
\newblock {\em bioRxiv}, 2023.

\bibitem{Yu2024ddaffinity}
Guanglei Yu, Qichang Zhao, Xuehua Bi, and Jianxin Wang.
\newblock {DDAffinity}: predicting the changes in binding affinity of multiple point mutations using protein 3d structure.
\newblock {\em Bioinformatics}, 40(Supplement\_1):i418--i427, 06 2024.

\bibitem{liu2024predicting}
Shiwei Liu, Tian Zhu, Milong Ren, Chungong Yu, Dongbo Bu, and Haicang Zhang.
\newblock Predicting mutational effects on protein-protein binding via a side-chain diffusion probabilistic model.
\newblock {\em Adv. Neural Inf. Process. Syst.}, 36, 2024.

\bibitem{stranges2013comparison}
P~Benjamin Stranges and Brian Kuhlman.
\newblock A comparison of successful and failed protein interface designs highlights the challenges of designing buried hydrogen bonds.
\newblock {\em Prot. Sci.}, 22(1):74--82, 2013.

\bibitem{buss2018foldx}
Oliver Bu{\ss}, Jens Rudat, and Katrin Ochsenreither.
\newblock {FoldX} as protein engineering tool: better than random based approaches?
\newblock {\em Comput. Struct. Biotechnol. J.}, 16:25--33, 2018.

\bibitem{Barlowflexddg}
Kyle~A. Barlow, Shane ^^c3^^93 Conch^^c3^^bair, Samuel Thompson, Pooja Suresh, James~E. Lucas, Markus Heinonen, and Tanja Kortemme.
\newblock {Flex ddG}: Rosetta ensemble-based estimation of changes in protein^^e2^^80^^93protein binding affinity upon mutation.
\newblock {\em J. Phys. Chem. B}, 122(21):5389--5399, 2018.
\newblock PMID: 29401388.

\bibitem{settles2009active}
Burr Settles.
\newblock Active learning literature survey.
\newblock Computer Sciences Technical Report 1648, University of Wisconsin-Madison, 2009.

\bibitem{reker2015active}
Daniel Reker and Gisbert Schneider.
\newblock Active-learning strategies in computer-assisted drug discovery.
\newblock {\em Drug Discov. Today}, 20(4):458--465, 2015.

\bibitem{reker2017active}
Daniel Reker, Petra Schneider, Gisbert Schneider, and J~B Brown.
\newblock Active learning for computational chemogenomics.
\newblock {\em Future Med. Chem.}, 9(4):381--402, 2017.

\bibitem{Wang2023-ht}
Lei Wang, Shao-Hua Shi, Hui Li, Xiang-Xiang Zeng, Su-You Liu, Zhao-Qian Liu, Ya-Feng Deng, Ai-Ping Lu, Ting-Jun Hou, and Dong-Sheng Cao.
\newblock Reducing false positive rate of docking-based virtual screening by active learning.
\newblock {\em Brief. Bioinform.}, 24(1):bbac626, 2023.

\bibitem{konze2019reaction}
Kyle~D Konze, Pieter~H Bos, Markus~K Dahlgren, Karl Leswing, Ivan Tubert-Brohman, Andrea Bortolato, Braxton Robbason, Robert Abel, and Sathesh Bhat.
\newblock Reaction-based enumeration, active learning, and free energy calculations to rapidly explore synthetically tractable chemical space and optimize potency of cyclin-dependent kinase 2 inhibitors.
\newblock {\em J. Chem. Inf. Model.}, 59(9):3782--3793, 2019.

\bibitem{loeffler2024optimal}
Hannes Loeffler, Shunzhou Wan, Marco Kl{\"a}hn, Agastya Bhati, and Peter Coveney.
\newblock Optimal molecular design: Generative active learning combining {REINVENT} with absolute binding free energy simulations.
\newblock {\em ChemRxiv}, 2024.

\bibitem{seo2022accelerating}
Seung-woo Seo, Min~Woo Kwak, Eunji Kang, Chaeun Kim, Eunyoung Park, Tae~Hyun Kang, and Jinhan Kim.
\newblock Accelerating antibody design with active learning.
\newblock {\em BioRxiv}, 2022.

\bibitem{lu2021k}
Tianchi Lu.
\newblock K-nearest robust active learning on big data and application in epitope prediction.
\newblock {\em Wirel. Commun. Mob. Comput.}, 2021(1):8752022, 2021.

\bibitem{gessner2024active}
Alexandra Gessner, Sebastian~W Ober, Owen Vickery, Dino Ogli{\'c}, and Talip U{\c{c}}ar.
\newblock Active learning for affinity prediction of antibodies.
\newblock {\em arXiv preprint arXiv:2406.07263}, 2024.

\bibitem{zhu2022large}
Fangqiang Zhu, Feliza~A Bourguet, William~FD Bennett, Edmond~Y Lau, Kathryn~T Arrildt, Brent~W Segelke, Adam~T Zemla, Thomas~A Desautels, and Daniel~M Faissol.
\newblock Large-scale application of free energy perturbation calculations for antibody design.
\newblock {\em Sci. Rep.}, 12(1):12489, 2022.

\bibitem{lu2024alphafold3}
Wei Lu, Jixian Zhang, Jihua Rao, Zhongyue Zhang, and Shuangjia Zheng.
\newblock {AlphaFold3}, a secret sauce for predicting mutational effects on protein-protein interactions.
\newblock {\em bioRxiv}, 2024.

\bibitem{jankauskaite2019skempi}
Justina Jankauskait{\.e}, Brian Jim{\'e}nez-Garc{\'\i}a, Justas Dapk{\=u}nas, Juan Fern{\'a}ndez-Recio, and Iain~H Moal.
\newblock {SKEMPI 2.0}: an updated benchmark of changes in protein--protein binding energy, kinetics and thermodynamics upon mutation.
\newblock {\em Bioinformatics}, 35(3):462--469, 2019.

\bibitem{olsen2022ablang}
Tobias~H Olsen, Iain~H Moal, and Charlotte~M Deane.
\newblock {AbLang}: an antibody language model for completing antibody sequences.
\newblock {\em Bioinform. Adv.}, 2(1):vbac046, 2022.

\bibitem{mcinnes2018umap}
Leland McInnes, John Healy, and James Melville.
\newblock Umap: Uniform manifold approximation and projection for dimension reduction.
\newblock {\em arXiv preprint arXiv:1802.03426}, 2018.

\end{thebibliography}
\clearpage


\end{document}